\documentclass[aps,prd,twocolumn,showpacs,draft,floatfix]{revtex4}
\begin{document}

\title{Alternative proposal to modified Newtonian dynamics}
\author{Juan~M.~Romero}
\email{sanpedro@nucleares.unam.mx}
\affiliation{Instituto de Ciencias Nucleares, UNAM, Apartado Postal 70-543, 
M\'exico 04510 DF, M\'exico}
\author{Adolfo~Zamora}
\email{zamora@nucleares.unam.mx}
\affiliation{Instituto de Ciencias Nucleares, UNAM, Apartado Postal 70-543, 
M\'exico 04510 DF, M\'exico}
\date{\today}

\begin{abstract}
From a study of conserved quantities of the so-called Modified 
Newtonian Dynamics (MOND) we propose an alternative to this theory. 
We show that this proposal is consistent with the Tully-Fisher law, 
has conserved quantities whose Newtonian limit are the energy and 
angular momentum, and can be useful to explain cosmic acceleration.
The dynamics obtained suggests that, when acceleration is very small, 
time depends on acceleration. This result is analogous to that of 
special relativity where time depends on velocity.
\end{abstract}


\maketitle
Nowadays there are various observational results in astrophysics whose 
explanation represents a challenge for theoretical physics. One of those 
problems is to explain the rotation curves of the galaxies. Observations 
indicate a relationship $V^{4}\propto M$ for the speed $V$ of the distant 
stars in a galaxy of mass $M.$ However, as the only force acting on those 
stars is gravity and their trajectories are circles, Newtonian dynamics 
indicates that the relationship to hold is $V^{2}=GM/r$, where $r$ is the 
distance from the star to the center of the galaxy. To account for the 
difference, some authors assume the existence of a sort of matter that 
does not radiate: the so-called dark matter. There are, however, other 
proposals which assume modifications to the gravitational field or to the 
laws of dynamics. By considering the behavior of the speed of the distant
stars, M.~Milgrom proposed a modification to Newton's second law as 
\cite{milgrom:gnus}
\begin{equation}
m\mu(z)\frac{d^2 x^{i}}{dt^{2}}=F^{i}, \quad i=1,2,3;
\label{eq:1}
\end{equation}  
where $z=|\ddot x|/a_{0}=\sqrt{\ddot x_{i} \ddot x^{i}}/a_{0}$,  
$a_{0}\approx 10^{-8}$cm/s$^{2}$ and $\mu(z)$ is a function satisfying
\begin{equation}
\mu(z)=\left\{
\begin{array}{ccc}
1 & \,\, {\rm if} & z \gg 1,\\
z & \,\, {\rm if} & z \ll 1.
\end{array}
\right.
\end{equation}
This proposal is usually called Modified Newtonian Dynamics (MOND). From 
it one can see that, in the MOND limit ($z\ll 1, \mu(z)=z$), a particle 
describing a circular trajectory in the potential $U=-GMm/r$ satisfies
\begin{equation}
V^{4}=a_{0}GM; \label{eq:tf}
\end{equation} 
which is consistent with the Tully-Fisher law: $L_{K}\propto V^{4}$,
where $L_{K}$ is the infrared luminosity of the disk galaxy 
\cite{tully:gnus}. Also interesting appears the fact that the constant 
$a_{0}$ can be written as
$a_{0} =cH_{0}/6\approx 10^{-8}$cm/s$^{2}$, with $H_{0}$ the Hubble 
constant and $c$ the speed of light; or alternatively by using the 
Eddington-Weinberg relation \cite{edw:gnus}, 
$\hbar^{2} H_{0}\approx Gcm_{N}^{3}$, as 
$a_{0}\approx m_{N}^{3}c(6m_{p}^{3}t_{p})^{-1}$, where $m_{N}$ is the 
proton mass and $m_{p}=(\hbar c/G)^{1/2}$ and $t_{p}=(\hbar G/c^{5})^{1/2}$ 
are the Planck mass and time respectively. This can be just a coincidence, 
but it could also indicate the existence of a fundamental relation between 
physics at very large and very small scales.\\

MOND is a purely phenomenological theory but it explains most of the
galaxy rotation curves without introducing dark matter \cite{sander:gnus}. 
Its simplicity is what makes it attractive. Extensions to MOND at the 
level of the gravitational field can be found in 
\cite{beke:gnus,soussa:gnus,beke1:gnus,sanders:gnus,lue:gnus}. 
Phenomenological implications of those can be seen in \cite{lue:gnus,esos:gnus}. 
But despite its achievements, MOND has problems of its own. A crucial one 
is the lack of conserved quantities as energy. In this work we perform a 
study of MOND's constants of motion and, by defining an energy, propose 
an equation of motion alternative to (\ref{eq:1}). This proposal has 
several conserved quantities that in the Newtonian limit ($z\gg 1, \mu(z)=1$) 
reduce to the usual ones: energy and angular momentum are two of them.
A generalization of the virial theorem is also provided. It is shown, in 
addition, that this proposal can be useful to explain cosmic acceleration. 
Finally, we show that a possible interpretation of the dynamics is that, for 
accelerations of the order of $a_{0}$, time depends on acceleration. This is 
analogous to special relativity where time depends on velocity.\\

Let us start by considering modified Newton's second law (\ref{eq:1}). By 
using spherical polar coordinates and assuming a central force field,
this equation can be written as
\begin{eqnarray}
m\mu(z)
\left(\ddot r-r\dot\theta^{2}-r\dot\phi^{2}\sin^{2}\theta \right)&=&
-\frac{\partial U}{\partial r}, \label{eq:polmo1}\\
m\mu(z)\left(r\ddot \theta +2\dot r\dot \theta- 
r\dot\phi^{2}\sin\theta\cos\theta\right)&=&0, \label{eq:polmo2}\\
m\mu(z)\frac{d}{dt}\left(r^{2}\dot\phi\sin^{2}\theta\right)&=&0. 
\label{eq:polmo3}
\end{eqnarray}
If $\mu(z)\not =0$, then $\theta=\pi/2$ is a solution to 
(\ref{eq:polmo2}); and Eq. (\ref{eq:polmo3}) implies that the quantity
\begin{equation}
L=r^{2}\dot\phi, \label{eq:angular}
\end{equation}
is conserved. By using these and $U=-GMm/r$, Eq. (\ref{eq:polmo1}) 
reduces to
\begin{equation}
\mu(z)
\left(\ddot r-\frac{L^{2}}{ r^{3}}\right)
=-\frac{GM}{r^{2}},\qquad  
z=\frac{1}{a_{0}}\bigg|\left(\ddot r-\frac{L^{2}}{ r^{3}}\right)\bigg|.
\end{equation}
In the MOND limit this equation becomes
\begin{equation}
\bigg|\left(\ddot r-\frac{L^{2}}{ r^{3}}\right)\bigg|
\left(\ddot r-\frac{L^{2}}{ r^{3}}\right)
=-\frac{a_{0}GM}{r^{2}};\label{eq:radmon}
\end{equation}
which implies the constraint
\begin{equation}
\left(\ddot r-\frac{L^{2}}{ r^{3}}\right)<0.\label{eq:asmon}
\end{equation}
By using this, Eq. (\ref{eq:radmon}) can be written as
\begin{equation}
\left(\ddot r-\frac{L^{2}}{ r^{3}}-\frac{\sqrt{a_{0}GM} }{r}
\right)\left(\ddot r-\frac{L^{2}}{ r^{3}}
+\frac{\sqrt{a_{0}GM} }{r}
\right)=0.
\end{equation}
This implies that the particle's trajectory must satisfy either
\begin{equation}
\ddot r-\frac{L^{2}}{ r^{3}}-\frac{\sqrt{a_{0}GM} }{r}=0,
\label{eq:as1}
\end{equation}
or
\begin{equation}
\ddot r-\frac{L^{2}}{ r^{3}}
+\frac{\sqrt{a_{0}GM} }{r}=0,\label{eq:as2}
\end{equation}
or both, but the constraint (\ref{eq:asmon}) is not compatible with
(\ref{eq:as1}) and therefore the whole equation (\ref{eq:radmon})
is reduced to (\ref{eq:as2}). Clearly, for Eq. (\ref{eq:as2})
the quantity
\begin{equation}
{\cal E}=\frac{\dot r^{2}}{2}+\frac{L^{2}}{2 r^{2}}+\sqrt{a_{0}GM} 
\ln r,
\label{eq:nermon}
\end{equation}
is conserved. This corresponds to the energy per unit mass of a particle 
moving in the potential $U(r)=\sqrt{a_{0}GM}\ln r$. It is tempting to 
take $\cal E$ as the energy of the system; however, this quantity is 
conserved only in the MOND limit and does not reduce to the usual energy 
in the Newtonian limit ($\mu(z)=1$). This makes it unsuitable.\\

Looking for alternatives, one can see that for a particle describing 
trajectories with $\dot z=0$ (circles are examples), the quantity
\begin{equation}
E=\frac{m\mu(z)}{2} \frac{dx^{i}}{dt}\frac{dx_{i}}{dt}+ U(x),
\end{equation}
is conserved. In fact,
\begin{equation}
\dot E=\left(m\frac{\mu\prime(z)\dot z}{2}\frac{dx^{i}}{dt}+m\mu(z)
\frac{d^{2}x^{i}}{dt^{2}}+\frac{\partial U(x)}{\partial x_{i}}\right)
\frac{dx_{i}}{dt}=0
\end{equation}
because of MOND equation (\ref{eq:1}). Here $\mu\prime(z)=d\mu(z)/dz$.
Notice that this quantity is conserved for every $\mu(z)$ and $U(x)$, 
and reduces to the usual energy in the Newtonian limit. In this sense it 
can be said that $E$ does provide a good definition of energy. Requesting 
conservation of this quantity, now for any trajectory, implies that the 
equation of motion
\begin{equation}
m\mu(z) \frac{d^{2} x^{i}}{dt^{2}}+ m\frac{\mu\prime(z)\dot z}{2} 
\frac{dx^{i}}{dt}=F^{i}
\label{eq:monp}
\end{equation}
must hold. Clearly, when $\dot z\approx 0$ this reduces to the modified 
Newton's second law (\ref{eq:1}) and is therefore consistent with the 
Tully-Fisher law.\\

Eqs. (\ref{eq:1}) and (\ref{eq:monp}) coincide in the Newtonian limit, 
but differ in any other case for non-circular trajectories. This is not
an issue as stars with the more non-circular trajectories are those close 
to the galaxy center; and they are outside the MOND regime. Distant stars,
on the other hand, are in the MOND regime and have trajectories that can 
be approximated by circles. Let us then see how Eq. (\ref{eq:monp}) differs 
from (\ref{eq:1}) for trajectories close to the circle. In general only 
magnitudes of velocity and acceleration of the distant stars can be measured, 
so it is appropriate to look at magnitude differences only. For Eq. 
(\ref{eq:1}), $|F|=m\mu(z)|\ddot x|$; but for (\ref{eq:monp}),
\begin{eqnarray}
|F|=m\mu(z)|\ddot x|
\sqrt{1+\frac{\mu^{\prime}(z) \dot z}{\mu(z)|\ddot x|^{2}}
\left(\ddot x\cdot \dot x + \dot x^{2}
\frac{\dot z\mu^{\prime}(z) }{4\mu(z)}\right)}.
\end{eqnarray}
Now, by assuming an elliptical trajectory:
$x^{i}=r_{0}(\cos\omega t, \sqrt{1-e^{2}}\sin\omega t, 0)$, with $e$ the
eccentricity; in the MOND limit and to the lowest order in $e$, one obtains
\begin{eqnarray}
|F|=m\mu(z)|\ddot x|\left(1-\frac{3}{32}e^{4}f(t)\right),
\label{eq:mag}
\end{eqnarray}
where $0\le f(t)\le 1$.
Thus, for the correction term to be $1\%$ of the magnitude 
$|F|=m\mu(z)|\ddot x|$, a large eccentricity $e\approx 0.57$ is required. 
In this sense Eq. (\ref{eq:monp}) is not so different from (\ref{eq:1}).\\

An advantage of (\ref{eq:monp}) over (\ref{eq:1}), though, is that in 
addition to energy it has several conserved quantities. For instance, 
for potentials $U$ depending on the distance $r$ only, Eq. (\ref{eq:monp}) 
implies conservation of the quantity
\begin{equation}
L_{i}=\epsilon_{ijk}x^{j}m\sqrt{\mu(z)} \frac{dx^{k}}{dt},
\end{equation}
which in the Newtonian limit reduces to angular momentum. If $U(r)=-GMm/r$,
also the quantity
\begin{eqnarray}
A_{i}=m\sqrt{\mu(z)} \epsilon_{ijk}\dot x_{j}L_{k}-\frac{GMm^{2}}{r} x_{i},
\end{eqnarray}
that in the Newtonian limit reduces to the Runge-Lenz vector, is 
conserved. In addition, it can be seen that for $U(r)=0$, the quantity
\begin{equation}
p_{i}=m\sqrt{\mu(z)} \frac{dx_{i}}{dt}, \label{eq:mome}
\end{equation}
is also conserved. This reduces to the usual momentum in the Newtonian 
limit.\\

Considering now $\big\langle\dot {\cal G}\big\rangle=
\lim_{T \to 0}\frac{1}{T}\int_{0}^{T}\dot {\cal G} dt=0$, where 
${\cal G}=p_{i}x^{i}$ with $p_{i}$ above, from Eq. (\ref{eq:monp}) we 
obtain
\begin{equation}
\big\langle\dot {\cal G}\big\rangle=
\left\langle\frac{F_{i}x^{i}}{\sqrt{\mu(z)}}\right\rangle+
\left\langle\sqrt{\mu(z)}m\dot x_{i}\dot x^{i}\right\rangle=0, 
\label{eq:virial}
\end{equation}
which is a generalization of the virial theorem \cite{landaumeca:gnus}.
For $U=-GMm/r$, and in the MOND limit, this equation yields
$\big\langle GMm/r\big\rangle\ll\big\langle m\dot x_{i}\dot x^{i}
\big\rangle$; which is qualitatively consistent with observations in 
galaxy clusters \cite{zwicky:gnus}.\\ 

Eq. (\ref{eq:monp}) is non-relativistic but from Newtonian cosmology one 
can still get some implications. It is worth noticing that Newtonian 
cosmology is an appropriate approximation when pressure can be neglected 
\cite{zeldovich:gnus}. Now, there are several ways to construct a Newtonian 
cosmology \cite{zeldovich:gnus,edw:gnus,milne:gnus} and, as all of them 
yield the same equations of motion, we take the simplest one. Let us assume 
the cosmological principle $x_{i}=R(t)\hat x_{i}(t_{0})$, with 
$\hat x_{i}(t_{0})$ a unit vector. Therefore a unit-mass particle in the 
gravitational field has energy
\begin{eqnarray}
E=\frac{1}{2}\mu(z)\dot R^{2}-\frac{GM}{R}, \qquad z=\frac{\ddot R}{a_{0}},
\end{eqnarray}
from where
\begin{eqnarray}
\mu(z)\frac{\dot R^{2}}{R^{2}}=-\frac{k}{R^{2}}+\frac{8\pi G}{3}\rho, 
\qquad \rho= \frac{3M}{4\pi R^{3}},
\label{eq:fried}
\end{eqnarray}
with $k=-E/2$. In the Newtonian limit ($\mu(z)=1$) this equation is
equivalent to Friedmann's for a pressureless-matter dominated universe.
In fact, if $E=0$ then $k=0$ and if $E\not =0$, $R$ can always be changed
to $\lambda R$ in such a way that $k$ only takes values $\pm 1$. Outside
the Newtonian limit Eq. (\ref{eq:fried}) is a MOND-like pressureless 
Friedmann equation. The $k=0$ case is particularly interesting as
recent observations indicate compatibility of the universe with this
value \cite{spergel:gnus}. For $k=0$ and in the Newtonian limit, the solution
to (\ref{eq:fried}) is of the form $R(t)\propto t^{2/3}$. In this case the 
deceleration parameter $q_{0}=-\ddot R R/\dot R^{2}>0$. Recent observations 
\cite{ob:gnus}, however, provide strong evidence of an accelerated 
universe with $q_{0}<0$. Now, by considering the MOND limit ($\mu(z)=z$) 
of Eq. (\ref{eq:fried}) one obtains
$\ddot R \dot R^{2}/R^{2}= 8\pi G a_{0}\rho/3$. From this,
$\dot R =\beta \left[\ln \left(R/R_{0}\right)\right]^{1/4}$, with
$\beta^{4}=8GMa_{0}$ and $R_{0}$ an integration constant. Therefore,
$\ddot R =\beta^{2}\left(4R[\ln(R/R_{0})]^{1/2}\right)^{-1}$.
Notice that to be within the MOND regime, $R_{0}<R$ must hold and therefore 
$q_{0}=-(4\ln(R/R_{0}))^{-1}<0$; which suggests that a relativistic 
generalization to the theory here presented could be useful to explain the 
universe acceleration without introducing dark energy.\\

The problem of structure formation can, in principle, also be tackled with 
Eq. (\ref{eq:fried}). However, from the equation of motion of the usual 
Newtonian cosmology at the structure formation epoch (SFE) one gets 
$|\ddot R/a_0|=|4\pi G\rho R/3|\approx 10^{8}$, which indicates that
Newton dynamics must not be replaced by MOND. Notice that if $a_0=cH_0/6$
is changed to $a_{0,{\scriptscriptstyle{\rm SFE}}}=cH_{\scriptscriptstyle{\rm SFE}}/6$, 
with $H_{\scriptscriptstyle{\rm SFE}}$ being the Hubble's constant at the SFE,
then $|\ddot R/a_{0,{\scriptscriptstyle{\rm SFE}}}|\approx 1$, and therefore 
it is necessary to consider MOND's corrections to Newton dynamics in this 
universe epoch. It is possible that a relativistic generalization to MOND 
may imply variation of $a_0$ with time so as to have implications in 
the SFE. Some of the properties a relativistic generalization to MOND must 
have can be found in \cite{soussa2:gnus}.\\

To interpret Eq. (\ref{eq:monp}) let us consider
\begin{equation}
m\frac{1}{\dot \tau^{2}} \frac{d^{2} x^{i}}{dt^{2}}-
m\frac{\ddot \tau }{\dot \tau^{3}} \frac{dx^{i}}{dt}=F^{i},
\qquad \dot \tau=\frac{d\tau}{dt}.
\label{eq:rep}
\end{equation}
Notice that if $\tau=t$, this equation reduces to Newton's second law.
Eq. (\ref{eq:rep}) is in fact a generalized Newton's second law where 
the time $\tau$ can depend on other variables. In particular, by taking
\begin{equation}
\frac{1}{\dot \tau^{2}}=\mu(z), \quad 
z=\frac{1}{a_{0}}\sqrt{\frac{d^{2} x_{i}}{dt^{2}}
\frac{d^{2} x^{i}}{dt^{2}}}, \label{eq:tim}
\end{equation}
Eq. (\ref{eq:rep}) equals (\ref{eq:monp}). Thus, Eq. (\ref{eq:monp})
can be interpreted as a Newton's second law where time depends on 
acceleration.\\

Another dynamics where time depends on other variables is the
relativistic one. Newton's second law in the relativistic case can be
written as \cite{landau:gnus}
\begin{equation}
m\frac{d^{2}x^{\alpha}}{d\tau^{2}}=
m\frac{1}{\dot \tau^{2}} \frac{d^{2} x^{\alpha}}{dt^{2}}-
m\frac{\ddot \tau }{\dot \tau^{3}} \frac{dx^{\alpha}}{dt}
=\frac{f^{\alpha}}{c}, \quad \alpha= 0,1,2,3; \label{eq:repre}
\end{equation}
where
\begin{equation}
\dot \tau=\gamma^{-1}, \qquad \gamma^{-1}=
\sqrt{1-\frac{\dot x^{i} \dot x_{i}}{c^{2}}}. \label{eq:tir}
\end{equation}
This provides analogies between the well known relativistic dynamics
and that given by Eq. (\ref{eq:monp}). Similarities between conserved 
quantities can be seen, for instance, by remembering that in special
relativity the conserved momentum is no longer $p_{i}=m\dot x_{i}$, 
but $p_{i}= m \gamma\dot x_{i}$ \cite{landau:gnus}; whereas for the
dynamics of (\ref{eq:monp}) is that from Eq. (\ref{eq:mome}). Finally,
it is straightforward to see that Eq. (\ref{eq:tir}) can be obtained 
from the line element
\begin{equation}
ds^{2}=c^{2}(dt)^{2}- dx^{i}dx_{i}=c^{2}(d\tau)^{2},
\end{equation}
whereas Eq. (\ref{eq:tim}) follows from
\begin{equation}
dS^{2}=a_{0}(dt)^{2}-\frac{(1-\mu^{-1}(z))}{a_{0}z^{2}} dv^{i}dv_{i}
=a_{0}(d\tau)^{2}.
\end{equation}
This suggests that a more general theory to the one here presented
may imply that, in addition to time, some geometrical quantities as 
distance also depend on acceleration.\\

To summarize, we have presented an alternative proposal to MOND which 
is consistent with the Tully-Fisher law and that has several conserved 
quantities whose Newtonian limit is the usual one. A generalization of 
the virial theorem is also provided. It is shown that this proposal is 
useful to explain cosmic acceleration. The dynamics obtained suggests 
that, for accelerations of the order of $a_{0}$, time depends on 
acceleration. It is worth mentioning that there are already proposals 
to tackle the problem of MOND's constants of motion by modifying 
Poisson's equation for the gravitational field \cite{beke:gnus}. Those 
conserved quantities are, however, not for the particle but for the 
gravitational field.\\

The authors acknowledge the ICN-UNAM for its kind hospitality and AZ 
also for financial support.

\end{document}